\documentclass{IEEEtran}

\usepackage[utf8]{inputenc}
\usepackage[newfloat]{minted}
\usepackage{quantikz}
\usepackage[colorlinks]{hyperref}
\usepackage{svg}
\usepackage{amsmath}
\usepackage{mathtools}

\usepackage{amssymb}
\usepackage{pifont}
\usepackage{graphicx}
\usepackage{algpseudocode}

\usepackage{algcompatible}
\usepackage{algpseudocode}
\usepackage{algorithm}
\usepackage{algorithmicx}

\algnewcommand\algorithmicreturn{\textbf{return}}
\algnewcommand\RETURN{\State \algorithmicreturn}%
\usepackage[justification=centering]{caption}
\usepackage{subcaption}
\usepackage{placeins}

\usepackage{subcaption}
\usepackage{xcolor}
\usepackage{array}
\newcolumntype{P}[1]{>{\centering\arraybackslash}p{#1}}
\newcolumntype{M}[1]{>{\centering\arraybackslash}m{#1}}
\newcolumntype{Q}[1]{>{\arraybackslash}m{#1}}
\usepackage{comment}
\usepackage{float}
\usepackage{makecell}
\newcommand{\linebreakand}{%
  \end{@IEEEauthorhalign}
  \hfill\mbox{}\par
  \mbox{}\hfill\begin{@IEEEauthorhalign}
}

\usepackage{tikz}
\usetikzlibrary{shapes.geometric, arrows, decorations.pathmorphing}
\tikzstyle{startstop} = [rectangle, rounded corners,text centered, draw=black, fill=red!30]
\tikzstyle{io} = [trapezium, trapezium left angle=70, trapezium right angle=110, minimum width=3cm, minimum height=1cm, text centered, draw=black, fill=blue!30]
\tikzstyle{process} = [rectangle, text centered, draw=black, fill=orange!30]

\tikzstyle{decision} = [diamond, text centered, draw=black, fill=green!30]
\tikzstyle{arrow} = [thick,->,>=stealth]

\newcommand{\vew}[1]{
	\edef\start{\the\pgfmatrixcurrentrow-\the\pgfmatrixcurrentcolumn}
	\edef\end{\the\numexpr#1+\pgfmatrixcurrentrow\relax-\the\pgfmatrixcurrentcolumn}
	\expandafter\expandafter\expandafter\vewexplicit\expandafter\expandafter\expandafter{\expandafter\start\expandafter}\expandafter{\end}
}
\newcommand{\vewexplicit}[2]{
	\arrow[from=#1,to=#2,arrows,decorate,decoration={snake,amplitude=1pt,segment length=6.5pt}] {}
}
\usepackage{listings}
\usepackage{xcolor}

\definecolor{codegreen}{rgb}{0,0.6,0}
\definecolor{codegray}{rgb}{0.5,0.5,0.5}
\definecolor{codepurple}{rgb}{0.58,0,0.82}
\definecolor{backcolour}{rgb}{0.95,0.95,0.92}

\lstdefinestyle{mystyle}{
    backgroundcolor=\color{backcolour},   
    commentstyle=\color{codegreen},
    keywordstyle=\color{magenta},
    numberstyle=\tiny\color{codegray},
    stringstyle=\color{codepurple},
    basicstyle=\ttfamily\footnotesize,
    breakatwhitespace=false,         
    breaklines=true,                 
    captionpos=b,                    
    keepspaces=true,                 
    numbers=left,                    
    numbersep=5pt,                  
    showspaces=false,                
    showstringspaces=false,
    showtabs=false,                  
    tabsize=2
}

\usepackage{adjustbox}
\usepackage{multirow}
\usepackage{amssymb}
\usepackage{pifont}
\usepackage{siunitx}

\begin{document}
\bstctlcite{IEEEexample:BSTcontrol}

\title{Automated Circuit Depth Reduction of Quantum Subroutines via Compilation}

\author{
    \IEEEauthorblockN{Folkert de Ronde}
    \IEEEauthorblockA{Quantum \& Computer Engineering \\
        Delft University of Technology \\
        Delft, The Netherlands \\
        f.w.m.deronde@tudelft.nl
    }
\and \\
\IEEEauthorblockN{Stephan Wong}
\IEEEauthorblockA{
Quantum \& Computer Engineering \\
Delft University of Technology \\
Delft, The Netherlands \\
j.s.s.m.wong@tudelft.nl}
\and \\
\IEEEauthorblockN{Sebastian Feld}
\IEEEauthorblockA{
Quantum \& Computer Engineering \\
Delft University of Technology \\
Delft, The Netherlands \\
s.feld@tudelft.nl}}

\maketitle
\thispagestyle{plain}

\begin{abstract}


Optimizing quantum circuits by reducing circuit depth is essential for improving the efficiency and scalability of quantum algorithms, particularly as quantum hardware continues to evolve. This can be achieved by restructuring quantum algorithms to allow more parallelism. A compiler is needed to automatically detect and apply these optimizations. In this work, we focus on the optimization of two fundamental quantum subroutines: GHZ state creation and CNOT/CZ chain decomposition. Traditional implementations of these subroutines suffer from linearly increasing circuit depth, which limits scalability. We propose a compiler-driven approach that automatically detects and optimizes these two fundamental quantum subroutines. 
Our approach reduces circuit depth through constant-depth GHZ state creation, constant depth CZ chain decomposition, and logarithmic depth recursive CNOT chain decomposition, which enhance parallel execution. Performance analysis of benchmarked algorithms shows significant reductions in depth. However, our solution also results in an increased gate count, which makes our optimization a trade-off. The gate count for the CNOT chains is doubled, where logarithmic depth reduction is achieved. The reduced circuit depth results in more efficient algorithms by reducing execution time.

\end{abstract}



\section{Introduction}

Quantum computing has the potential to revolutionize computation by offering exponential speedups in fields such as chemistry, optimization, and cryptography. However, efficient quantum circuit design remains a critical challenge due to restricted qubit connectivity, gate overhead induced by compilation, and hardware noise constraints. As quantum algorithms grow in complexity (number of qubits and gates, but also circuit depth), optimizing quantum circuits becomes essential to increase circuit fidelity, minimize resource usage and reduce circuit depth. The reduction in circuit depth is achieved by improving the number of gates that can be executed in parallel. 

One key area of optimization lies in the decomposition of subroutines that exist within quantum algorithms. Two of such subroutines are GHZ state creation \cite{greenberger1989going} and CNOT/CZ chain operations \cite{PhysRevA.98.022322, sim2019expressibility}. These structures appear in quantum algorithms \cite{yimsiriwattana2004distributed, Kiss_2022}, including those used in entanglement generation, error correction \cite{ghosh2018automated}, variational quantum algorithms \cite{sim2019expressibility}, and quantum machine learning \cite{mansky2025solving}. Multiple implementations of the GHZ state creation have been proposed with difference in circuit depth, ranging from linear to constant. The conventional implementation of the CNOT/CZ chain routines suffers from limited parallelism: every subsequent gate is dependent on the previous gate, leading to linearly increasing circuit depth as the number of qubits grows. This inefficiency can significantly impact the circuit depth of a quantum circuit. A larger circuit depth results in longer execution time of a quantum algorithm, making them impractical for near-term quantum devices. The circuit depth in CNOT/CZ chains can be reduced by transforming the subroutine. The manner in which that is done will be explained in the body of the paper.

To address the issue of limited parallelism in both subroutines, we explore the following research question:
How can we identify two inherently sequential quantum subroutines, namely GHZ state creation and CNOT/CZ chains, and automate the detection and transformation for circuit depth reduction?

In this paper, we propose a transformation for CNOT and CZ chains, together with a compiler that can automatically detect and transform GHZ state creation and CNOT/CZ chain subroutines into their improved counterparts. 
In this paper, we propose a transformation for CNOT and CZ chains, together with a compiler that can automatically detect and transform GHZ state creation and CNOT/CZ chain subroutines into their improved counterparts.
For GHZ state generation, we use existent alternative circuits that reduce circuit depth to logarithmic or even constant depth while preserving logical correctness. For CNOT chains, we introduce recursive decomposition strategies that enable parallel execution on quantum hardware, reducing the overall circuit depth while preserving logical correctness. For CZ chains, we introduce a decomposition that reduces the circuit depth from any chain length to constant depth 2. Our compiler is capable of automatically detecting an decomposing these structures.

In addition to structural transformations, our compiler incorporates user-defined preferences, allowing trade-offs between gate count, circuit depth, and robustness against measurement noise. This flexibility enables tailored optimization strategies that adapt to the characteristics of both, algorithm and target hardware.

Our work has the following key contributions:
\begin{enumerate}
    \item  Decompositions of GHZ state creation with support for both parallel and noise-robust variants while maintaining logical structure.
    \item Transformations of CNOT chains to logarithmic depth while maintain logical structure.

    \item Transformation of CZ chains to constant depth of 2 while maintaining logical structure.

    \item A configurable compiler that supports preference-based optimization and customizable ordering of transformation passes.
\end{enumerate}

We evaluate our approach by analyzing circuit depth and gate count of circuits, before and after applying the transformations. We also demonstrate the effectiveness of these transformations through both, standalone benchmarks and integration into the Variational Quantum Eigensolver (VQE) algorithm. 

The remainder of this paper is structured as follows: Section \ref{sec:related_work} discusses other compilers that tackle gate count and hardware connectivity. Section \ref{sec:Algorithms} discusses the optimization of GHZ state creation and CNOT/CZ chain optimization. Section \ref{sec:Compiler_design} provides an overview of the compiler's detection and transformation strategies. Section \ref{sec:analysis} presents an analysis of the improvements achieved by our approach. Section \ref{sec:conclusion} summarizes our findings and outlines potential future research.

\section{Related work} \label{sec:related_work}

Circuit optimization has long been a central theme in quantum computing research, given the constraints of noisy intermediate-scale quantum (NISQ) devices. The application of optimizations to a quantum circuit is not trivial, which can be solved by using quantum compilers to apply these optimizations. Other works that use compilers to perform circuit optimizations focused on gate-level transformations and hardware mapping are\cite{amy2013meetinthemiddle, shaik2024optimal, chen2023nearest,bataille2022quantum, guo2025efficient}.  These efforts primarily reduce gate counts or ensure hardware connectivity, but do not target depth reductions in highly sequential parts of algorithms. Focusing on depth reduction allows for lower execution times of algorithms.  

The creation of GHZ states is well studied, with applications in entanglement distribution, and quantum error correction. The standard approach employs a chain of CNOT gates, leading to depth that scales linearly with the number of qubits \cite{greenberger1989going, nielsen2010quantum}. Even though the chain looks inevitable to overcome and an issue when hardware starts to scale up, it is possible to perform GHZ state creation in a more parallel fashion. Optimized strategies based on tree-like decompositions or parallel entanglement have been proposed \cite{cruz2019efficient, mooney2021generation, quek2024multivariate}. These methods reduce depth to logarithmic or even constant levels at the expense of additional measurements or ancilla qubits. Our work builds on these insights, but integrates them into a compiler framework, enabling automated detection and substitution of GHZ routines.

Entangling chains of gates, including CNOT and CZ sequences, appear in many variational ansätze such as the Variational Quantum Eigensolver (VQE) \cite{sim2019expressibility}. Previous studies have shown that naive implementations suffer from limited parallelism, resulting in linearly increasing depth. Our compiler extends these approaches by systematically recognizing CNOT chains, transforming the circuit, and ensuring constant-depth decomposition for CZ subroutines.

Quantum compilers such as Qiskit \cite{qiskit2024}, t|ket$\rangle$ \cite{sivarajah2020t}, and Quilc \cite{smith2016practical} include various optimization passes to reduce gate count, circuit depth, and error propagation. Many of these compilers focus on hardware-specific mapping and transpilation \cite{zulehner2019efficient}, while others explore algebraic simplification and optimizations based on commutativity \cite{amy2014polynomial, nam2018automated}. However, the depth reduction of structured subroutines has received less attention. Our work is complementary to these compilers: rather than replacing existing compiler functionalities, we propose a solution that specifically targets GHZ state creation and entangling chains, after these optimizations other compilers can still be used to map the resulting logical structure to hardware executable code. This strategy achieves significant depth savings with only modest gate overhead.

Previous work has underscored the critical role of circuit depth optimization in quantum algorithms. Although efficient GHZ state creation has been investigated, these efforts have largely remained isolated from compiler infrastructures. Our work closes this gap by creating automated detection and configurable transformations, and implementing them  directly into a compiler framework, bringing practical usability to previously standalone strategies.

In addition to the focus on GHZ state creation, we also introduce novel depth-reduction techniques for CZ and CNOT chains. Here again, the compiler plays a central role: it not only identifies opportunities for optimization automatically, but also applies them seamlessly, ensuring that advanced strategies translate into tangible improvements in real circuits.

\section{Fundamental subroutines: GHZ state creation and entanglement chains} \label{sec:Algorithms}
In this section, we discuss how GHZ creation and entangling chains can be structured in manners that have different advantages such as reduced circuit depth, gate count or measurements. The fundamentally used subroutines are the Greenberger-Horne-Zeilinger (GHZ)~\cite{greenberger1989going} state creation algorithm and the CNOT chain~\cite{sim2019expressibility} routine. While both routines have a straightforward decomposition, we propose an optimized decomposition strategy that reduces circuit depth. 

\subsection{GHZ State Creation}
\begin{figure}
    \centering
    \includegraphics[width=0.65\linewidth]{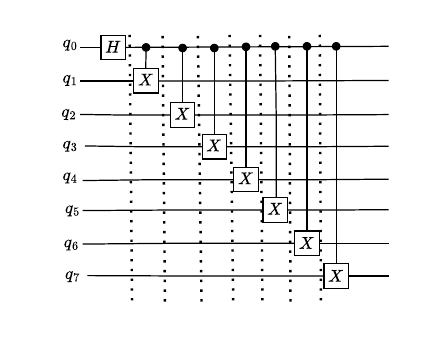}
    \caption{Circuit representation of a standard GHZ state generation in a configuration linear to the number of qubits. Gates in between dotted lines can be executed in parallel. }
    \label{fig:ghz_state_creation}
\end{figure}

\begin{figure}
    \centering
    \includegraphics[width=0.65\linewidth]{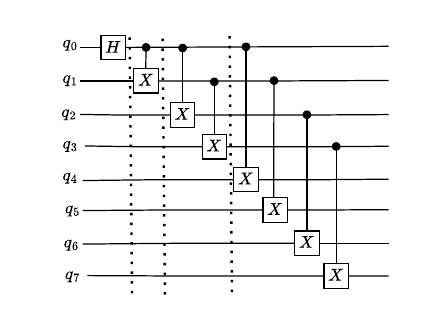}
    \caption{Circuit representation of an improved GHZ state generation scheme with depth logarithmic in the number of qubits. Gates in between dotted lines can be executed in parallel. }
    \label{fig:ghz_state_creation_improved}
\end{figure}

\begin{figure}
    \centering
    \includegraphics[width=0.65\linewidth]{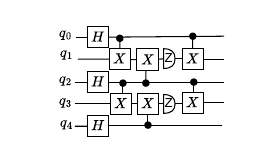}
    \caption{Circuit representation of an improved GHZ state generation scheme with constant depth. The design allows gates to be added fully in parallel as the number of qubits increases. The Z symbol denotes measurement in the Z basis.}
    \label{fig:ghz_state_creation_optimized}
\end{figure}

Multiple methods have been developed for GHZ state creation, as referenced in \cite{cruz2019efficient} and \cite{quek2024multivariate}. The standard approach is illustrated in Fig. \ref{fig:ghz_state_creation}, which has a circuit depth linear in the number of qubits. A solution with logarithmic circuit depth is depicted in Fig. \ref{fig:ghz_state_creation_improved}, while the implementation with constant circuit depth but additional measurements is depicted in Fig. \ref{fig:ghz_state_creation_optimized}.

The first two implementations have limited parallelism capabilities, which restricts their efficiency in quantum computations when executed on quantum computers. Optimizing GHZ state creation by changing the circuit's structure can significantly improve the execution of quantum algorithms by reducing circuit depth and execution time.

\subsection{Chain subroutines}
Multiple types of chain subroutines appear in quantum circuits. In this section, we focus on three types of these chains, namely: a forward CNOT chain, a reversed CNOT chain and a CZ chain. All these types of chains can be restructured to reduce circuit depth. Therefore, to improve parallelism and reduce depth, we introduce targeted decomposition strategies for each of the chain types. Each strategy considers the directionality and commutation properties of the gates involved to enable parallel execution.

A forward CNOT chain is a sequence of controlled-NOT (\texttt{CX}) gates arranged in ascending qubit order (e.g., \texttt{CX($q_0$, $q_1$)}, \texttt{CX($q_1$, $q_2$)}, ...). These structures are commonly used to entangle qubits.

\begin{figure*}[!tbp]
  \centering
  \begin{subfigure}[b]{0.25\textwidth}
    \includegraphics[width=\textwidth]{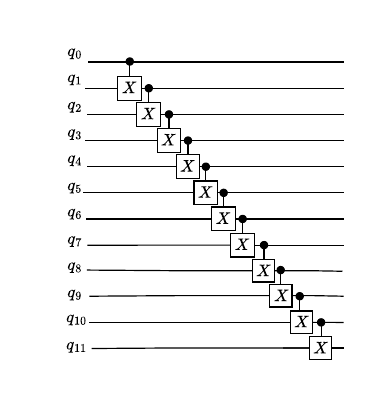}
    \caption{}\label{fig:cnot_chain_common}
  \end{subfigure}
  \hfill
  \begin{subfigure}[b]{0.31\textwidth}
    \includegraphics[width=\textwidth]{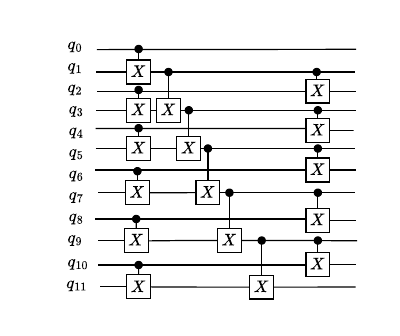}
    \caption{}
    \label{fig:cnot_chain_iterative}
  \end{subfigure}
  \hfill
  \begin{subfigure}[b]{0.31\textwidth}
    \includegraphics[width=\textwidth]{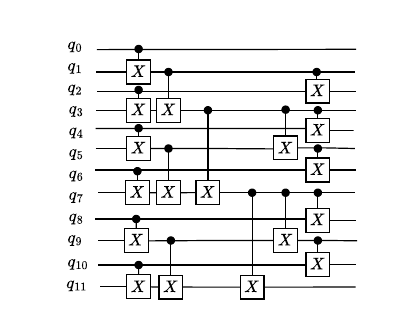}
    \caption{}
    \label{fig:cnot_chain_recursion}

  \end{subfigure}
  \caption{ 
  (a) Circuit representation of a standard forward CNOT chain, exhibiting linear depth in the number of involved qubits.
(b) Representation of an improved CNOT chain circuit by using the gate property shown in Fig. \ref{fig:gate_property} that reduces the circuit depth by half.
(c) Circuit representation of the recursive application of the method shown in (b), repeated until no further improvement is possible. The recursive approach achieves a circuit depth logarithmic in the number of qubits.}
\end{figure*}

The standard implementation, depicted in Fig. \ref{fig:cnot_chain_common}, follows a strictly sequential pattern, where each gate has a dependency on the gate before it, preventing any parallel execution. This leads to a circuit depth that scales linearly with the number of qubits involved.

\begin{figure*}[!tbp]
  \centering
  \begin{subfigure}[b]{0.25\textwidth}
    \includegraphics[width=\textwidth]{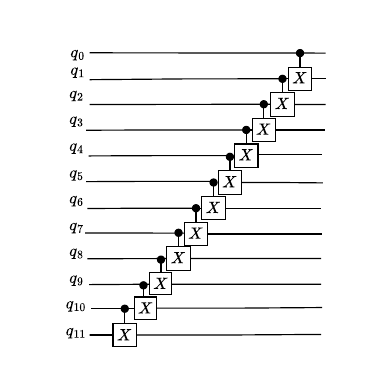}
    \caption{}\label{fig:reverse_cnot_chain}
  \end{subfigure}
  \hfill
  \begin{subfigure}[b]{0.31\textwidth}
    \includegraphics[width=\textwidth]{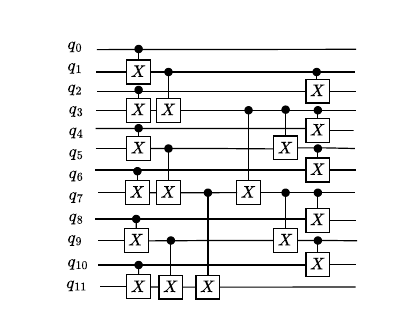}
    \caption{}
    \label{fig:reverse_cnot_chain_optimized}
  \end{subfigure}
  \hfill
  \begin{subfigure}[b]{0.25\textwidth}
    \raisebox{1.5cm}{\includegraphics[width=\textwidth]{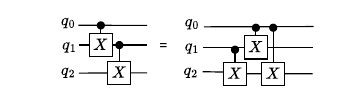}}
    \caption{}
    \label{fig:gate_property}

  \end{subfigure}
  \caption{ (a) Circuit representation of a standard reversed CNOT chain, exhibiting linear depth in the number of qubits.
(b) Recursive implementation of the reversed CNOT chain, where the optimization process is applied iteratively until no further improvement is possible. This results in logarithmic depth.
(c) Gate transformation rule used to obtain the improved CNOT chain solution.}
\end{figure*}

To reduce circuit depth, these CNOT chains can be arranged in a different manner, by using the transformation rule depicted in Fig. \ref{fig:gate_property}. The resulting solution is depicted in Fig. \ref{fig:cnot_chain_iterative}. In this case the circuit depth is reduced to half its original depth. In the figure it can be seen that a new CNOT chain is formed in the middle of the quantum circuit, therefore further gains can be achieved via recursive application of the same principle resulting in logarithmic depth. 
 The recursive approach is visualized in Fig.~\ref{fig:cnot_chain_recursion}. This solution provides circuit depth reduction for any chain longer than 5 CNOT gates. 
 It is important to note that even- and odd-length chains decompose slightly differently. Specifically, even-length chains introduce one additional \texttt{CNOT} gate that cannot be parallelized, slightly increasing the gate count.


Reverse CNOT chains\footnote{During the writing of this paper the authors have found another paper (\cite{remaud2025ancilla})that also shows the same optimization for reverse CNOT chains.} follow the same logical pattern as forward chains, but with descending qubit indices (e.g., \texttt{CX($N$, $N-1$)}, \texttt{CX($N-1$, $N-2$)}, ...), as depicted in Fig.~\ref{fig:reverse_cnot_chain}. The decomposition strategy mirrors that of forward chains, but applies the parallelization and recursion in reverse order, from high to low qubit indices. When the recursive optimization is applied, the resulting structure is depicted in Fig.~\ref{fig:reverse_cnot_chain_optimized}. As with forward chains, an even-length adjustment may be needed to maintain logical equivalence.

\begin{figure*}[!tbp]
  \centering
  \begin{subfigure}[b]{0.28\textwidth}
    \includegraphics[width=\textwidth]{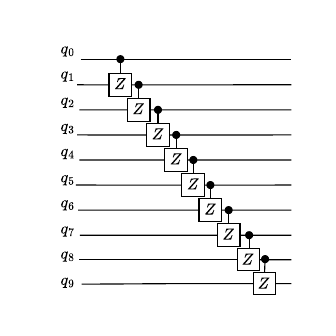}
    \caption{}\label{fig:cz_chain}
  \end{subfigure}
  \hfill
  \begin{subfigure}[b]{0.17\textwidth}
    \includegraphics[width=\textwidth]{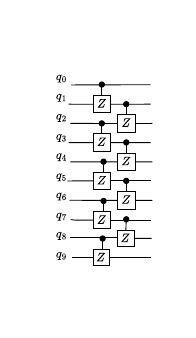}
    \caption{}
    \label{fig:cz_chain_optimized}
  \end{subfigure}
  \hfill
  \begin{subfigure}[b]{0.22\textwidth}
    \includegraphics[width=\textwidth]{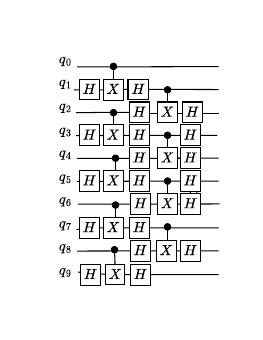}
    \caption{}
    \label{fig:cz_chain_decomposition}

  \end{subfigure}
  \begin{subfigure}[b]{0.22\textwidth}
    \includegraphics[width=\textwidth]{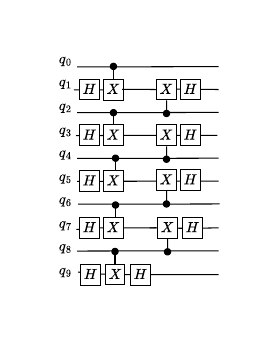}
    \caption{}
    \label{fig:cz_chain_optimized_decomposition}

  \end{subfigure}
  \caption{ (a) Circuit representation of a standard CZ chain. The solution shows linear depth in the number of involved qubits. (b) Improved circuit representation of a reversed CZ chain. The solution shows constant depth. (c) Circuit decomposition representation of the improved CZ chain when decomposing into CX gates. The solution shows constant depth. (d) Improved circuit decomposition representation of the improved CZ chain when decomposing into CX gates. The solution shows constant depth with additional depth reduction due to gate cancellation in the middle. }
\end{figure*}

A controlled-z (\texttt{CZ)} chain is depicted in Fig. \ref{fig:cz_chain}.  Unlike CX, the CZ gate is symmetric with respect to control and target qubits, which provides higher flexibility in parallel execution. The decomposition groups CZ gates into layers that can be executed in parallel. Fig.~\ref{fig:cz_chain_optimized} illustrates an optimized CZ chain, where gates are grouped into parallel layers with minimal impact on the overall circuit depth.

Beyond optimizing native CZ gate chains, it might also be necessary to decompose CZ gates into CNOT gates, for example when targeting quantum hardware without direct CZ support. Not all decompositions are equally efficient, for instance, a naive strategy would result in unnecessary depth and gate count. Fig. \ref{fig:cz_chain_decomposition} illustrates such a non-optimal decomposition of an already optimized CZ chain into CNOT gates. In contrast, Fig. \ref{fig:cz_chain_optimized_decomposition} shows an improved strategy where the CZ chain is decomposed with awareness of parallelization opportunities and gate cancellation rules, resulting in a more depth-efficient and hardware-conscious layout. This decomposition of CZ chains ensures that, even when translation into CNOT gates is required, the benefits of the original parallel structure are retained as much as possible.

Overall, these CNOT and CZ chain transformations are foundational to our compiler’s ability to restructure quantum circuits. Their recursive, symmetry-aware, and hardware-conscious design allows for depth reduction in both, synthetic benchmarks and practical quantum algorithms such as VQE.




\section{Automated detection and implementation of improved subroutines} \label{sec:Compiler_design}

The optimization of the previously presented quantum circuits is only one part of the solution. It is not always trivial to apply these transformations manually, and therefore an automated implementation of the transformation is needed. In the following, we present compiler passes that apply strategies for identification and automatic transformation of these subroutines into their improved counterparts. 
Specifically, this compiler is designed to identify GHZ state creation and CNOT/CZ chain subroutines within a quantum circuit and replace them with more efficient versions. 

The general workflow consists of four steps. (1) Identification of subroutines in a quantum circuit, (2) Identification of qubits within subroutine, (3) Transformation of the subroutine to an optimized version, (4) Return the reconstructed subroutine to its original location in quantum circuit.

The remainder of this section is structured as follows. First, we explain the identification and transformation used for GHZ state creation. Second, we focus on (CNOT/CZ) entanglement chains. Third, we will introduce a detection mechanism to prevent certain scenarios in which our optimizations might result in increased circuit depth. Finally, we discuss the speedup of our detection algorithm.

\subsection{GHZ State Creation}
A GHZ state subroutine can be detected by going through a quantum circuit and identifying where a Hadamard gate is happening. Once a Hadamard gate is found, the subsequent instructions are checked to be CNOT gates where they are either in a chain like fashion or with a single control qubit and multiple target qubits. All of the qubits in this interaction are then stored to be used for decomposition.  

Once a GHZ subroutine is detected, the compiler automatically decomposes it into the most efficient implementation depending on circuit size and user-defined preferences.

If the user prioritizes robustness against measurement noise, the compiler reconstructs the circuit using the structure shown in Fig.~\ref{fig:ghz_state_creation_improved}. This approach applies a logarithmically structured sequence of CNOT operations to reduce the circuit depth. The transformation proceeds as follows:

\begin{enumerate}
    \item Apply a Hadamard gate to qubit \( q_0 \).
    \item For each depth level (identified by the dotted lines in Fig.~\ref{fig:ghz_state_creation_improved}) \( d \in \{0, \ldots, \log_2(n) - 1\} \):
    \begin{itemize}
        \item For each qubit \( q_i \) used in the previous depth level, apply a CNOT from \( q_i \) to \( q_{i + 2^d} \), if \( q_{i + 2^d} \) exists.
    \end{itemize}
\end{enumerate}


Alternatively, if instead the user prefers maximal parallelism, the compiler transforms the circuit into the form shown in Fig.~\ref{fig:ghz_state_creation_optimized}. This version achieves constant depth through extensive parallelization, at the cost of introducing additional measurement operations. The transformation proceeds as follows:

\begin{enumerate}
    \item Apply Hadamard gates to all odd-indexed qubits.
    \item In the first entanglement layer, apply CNOT gates in parallel:
    \begin{itemize}
        \item From each odd-indexed qubit to its adjacent even-indexed upper neighbor (if it exists),
        \item From each odd-indexed qubit to its adjacent even-indexed lower neighbor (if it exists).
    \end{itemize}
    \item Measure all odd-indexed qubits.
    \item In the second entanglement layer, apply CNOT gates:
    \begin{itemize}
        \item From each measured odd-indexed qubit to its lower even-indexed neighbor (if applicable).
    \end{itemize}
\end{enumerate}

With these two approaches we can transform our GHZ state creation subroutine by either minimizing circuit depth or the amount of measurements, which results in a trade-off. Minimizing circuit depth results in the lowest execution time due to which the effects of depolarization and decoherence noise will be reduced. Reducing the amount of measurements results in a lower amount of measurement noise.

\subsection{Chain Subroutines} 
In quantum circuits, certain repeated structures, such as sequences of \texttt{CX} (CNOT) or \texttt{CZ} gates, can be grouped into chains and decomposed for better parallelism and reduced circuit depth. Although these two chains share a similar logical structure, each requires a distinct detection and decomposition strategy based on their operational semantics. 

To optimize such chains, our compiler includes a robust detection and decomposition system, capable of identifying and transforming both \textit{forward} and \textit{reverse} CNOT chains, as well as CZ chains. Even when additional, non-chain-breaking gates are interleaved within the sequence, the compiler is able to perform this task. 

\subsubsection{Chain Identification}
\begin{figure*}
    \centering
    \includegraphics[width = \linewidth]{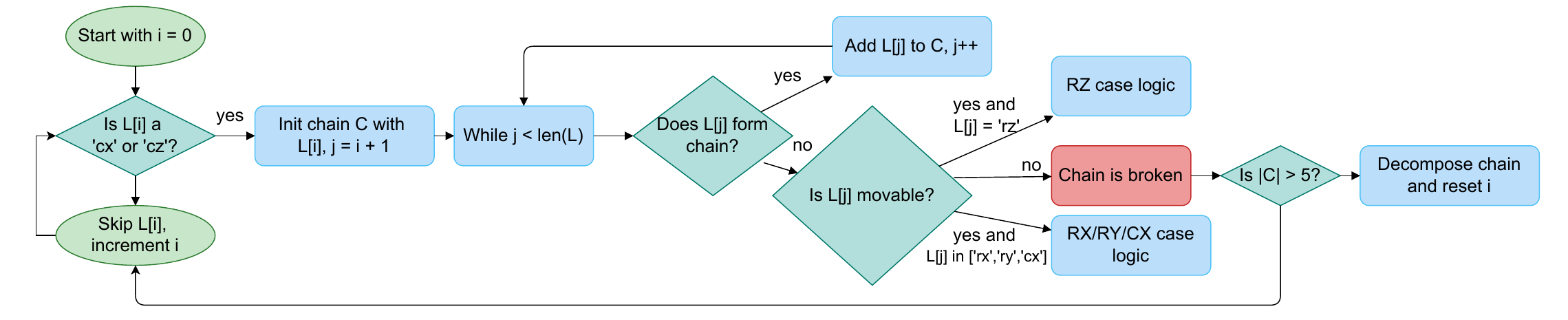}
    \caption{A flowchart describing the functionality of the detection algorithm. The algorithm is capable of detecting forward/backward CX and CZ chains. A chain is broken when an instruction is put in the chain that cannot be moved away.
    Decomposition is performed afterwards on the detected chain.}
    \label{fig:flowchart}
\end{figure*}

The compiler implements a detection algorithm that scans the quantum circuit instruction list $L$ to identify chains. This logic, shown as pseudo code in Alg. \ref{alg:detectChainsImproved} and as a flowchart in Fig. \ref{fig:flowchart}, enables the compiler to detect complex chains, i.e., even those that are not clearly structured. For example, a valid \texttt{CX} chain can still be recognized even when interleaved with gates in the qasm file. This could either be gates that can be commuted out of the chain, or are not dependent on the qubits in the chain at all. 
This includes non-dependent or commuting \texttt{rz}, \texttt{rx}, or \texttt{ry} gates, as long as they do not interfere with the qubits involved in the CNOT chain.

\begin{figure}
    \centering
    \includegraphics[width=0.5\linewidth]{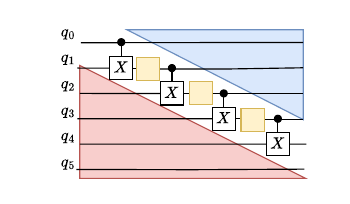}
    \caption{A CNOT chain with additional regions identified for other operations that can be moved. Operations in the blue region can be moved backwards in the circuit. Operations in the red region can be moved forward. Operations in the yellow box break the chain, except for RZ gates that can move through control qubits.}
    \label{fig:cnot_chain_to_identify}
\end{figure}

These single-qubit gates might \textit{commute} with \texttt{CX} operations under specific conditions. To help explain when these conditions are met, regions have been marked in Fig. \ref{fig:cnot_chain_to_identify}. In particular:

\begin{enumerate}
\item Yellow-box rules:
\begin{itemize}
\item If the operation is \texttt{RZ} on the \textit{control} qubit, mark it as \textit{movable backward (to the right)}.
\item If the operation is \texttt{RY} or \texttt{RX}, the chain is broken.
\item If the operation is a \texttt{CX} with its control qubit in the yellow region, it is not part of the chain if its target qubit coincides with any earlier control qubit.
\end{itemize}
\item Blue-region rule (backward-shift operations):  
\begin{itemize}  
    \item If the current operation acts on qubits that \textit{overlap} with the qubits in the chain, it must be pushed \textit{backward}.  
\end{itemize}  

\item Red-region rule (forward-shift operations):  
\begin{itemize}  
    \item If the current operation acts on qubits that do \textit{not overlap} with the qubits in the chain, it must be pushed \textit{forward}.  
\end{itemize}  
\end{enumerate}

The rules for the yellow boxes ensure that gate commutativity is upheld correctly. Z rotations can freely commute through control qubits and can therefore be moved backward. X and Y rotations cannot commute and therefore break the chain. CX gates can be part of the chain, or break the chain. A CX gate can break the chain if the next instruction has a target qubit where the control qubit should be, because this would enforce cyclic behavior. 

The rules for the blue and red boxes only shift operations to the front or the back of a circuit, based on whether they are already in front or in the back of the chain. This can be identified by determining whether any of the qubits that are currently under test are already part of the chain or not. 

By leveraging these commutation rules, the detection algorithm tolerates such intermediate gates and does not prematurely break chains, preserving the opportunity for optimization.

In order to deal with CZ and reversed CNOT chains, the aforementioned rules need slight modifications. In the case of a CZ chain, RZ operations can be moved backward or forward due to the commutative nature. In the case of a reverse CNOT chain, the RZ gate can be moved forward into the circuit instead of backward because it can only commute with the control gate of the chain.

\subsubsection{Detection of Intertwined Chains} 
Our detection algorithm has the ability to identify and separate multiple intertwined CNOT chains within the same region of the circuit. This scenario commonly arises in entanglement-heavy subroutines such as stabilizer preparation, where several chains may overlap. 

\begin{figure}
    \centering
    \includegraphics[width=0.5\linewidth]{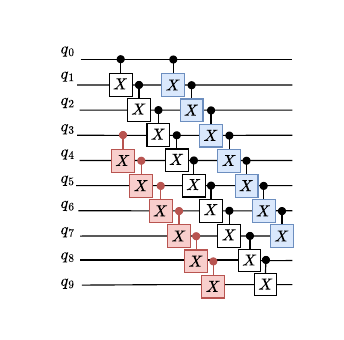}
    \caption{A representation of three CNOT chains that have intertwined operations. The compiler will see parts of all the chains at the same time and it has to be capable of identifying all chains. The identified red chain will be pushed forward while the blue chain is pushed backward. After the compiler has identified the middle chain, it will start searching from the start of the chain again.}
    \label{fig:compiler_cnot_chain_detection}
\end{figure}

As shown in Fig.~\ref{fig:compiler_cnot_chain_detection}, three distinct \texttt{CX} chains are present, marked in red, white, and blue. The white chain is detected first during the initial pass. However, upon completing the decomposition of this chain, the compiler does not simply continue from the end of the white circuit, because it would then not be able to identify the other two chains. Instead, it resets the iterator to the saved location of the first gate in the original chain detection, while the initial chain is already decomposed. This ensures that the compiler will then find the red chain next, after which the iterator is reset to the start of the red chain. At last the blue chain will then be found.


In this way:
\begin{itemize}
   
    \item The compiler stores the location of the first detected \texttt{CX} gate, and resets the iterator to this point after every chain transformation, enabling iterative discovery of all chains.
    
    \item This guarantees that gates moved leftward during decomposition (such as the red chain gates) are still detected and optimized appropriately.
    
    \item The blue chain, occurring after the white chain, is detected and optimized naturally in a later pass.
\end{itemize}

\subsubsection{Decomposition Strategies}

Once a chain is identified, the compiler applies a custom decomposition strategy based on the type and direction of the chain.

The decomposition for the forward CNOT chain is as follows:
\begin{enumerate}
    \item Apply \texttt{CX} gates from odd-indexed control qubits to their adjacent targets.
    \item Store control gates for potential recursive decomposition.
    \item Decompose the remaining chain recursively. 
    \item Stop recursion when fewer than five qubits remain, as deeper decomposition does not reduce circuit depth.
    \item Apply \texttt{CX} gates from even-indexed control qubits to their adjacent targets, based on the previously stored control gates.
    \item If the chain has even length, insert one additional \texttt{CX} at the end of the circuit to preserve logical equivalence.
\end{enumerate}

The decomposition for a reverse CNOT chain is as follows:

\begin{itemize}
    \item The decomposition mirrors the forward logic, but processes from high to low indices.
    \item Parallelization and recursive steps are applied in reverse order.
    \item Even length adjustments are also made at the end of the sequence.
\end{itemize}

The decomposition of CZ chains is as follows:
    Since \texttt{CZ} gates are symmetric (control and target are interchangeable), the decomposition always results in a set of instructions with constant depth of 2 independent of the CZ chain length.


\subsection{Improvement Detection}
\label{sec:improvement-detection}

A key design choice to focus on in our compiler is that it never degrades circuit depth. The compiler incorporates a mechanism that only applies a transformation if it results in an actual depth reduction. This ensures that the optimized circuit is always at least as good as the original in terms of depth.

The improvement detection process is shown in Alg. \ref{alg:improvement-detection} and works as follows. First, the compiler computes the circuit depth prior to a transformation using a depth analysis algorithm. To limit overhead, depth evaluation is restricted to CNOT chain that is under test and the subsequent 100 operations. After planning the transformation the depth is recalculated, and the results are stored along with the modified subroutine. If multiple CNOT chain transformations are possible (e.g., in the case of consecutive CNOT chains), the compiler evaluates the depth of each CNOT chain iteratively and compares their total effect on circuit depth. Among these options, only the configuration resulting in the lowest overall depth is applied. Once the transformation is performed, the corresponding CNOT chains are removed from the stored CNOT chain list. The detection process restarts whenever a new CNOT chain is encountered.

For users seeking more aggressive optimization, an optional parameter allows the compiler to always apply transformations, regardless of immediate depth reduction. While this setting may, in some cases, introduce temporary increases in depth, it can also enable further reductions, thus exploring a larger optimization space. This trade-off provides flexibility between conservative and exploratory optimization strategies.

\begin{algorithm}[t]
\caption{Improvement-Aware Transformation of Chains}
\label{alg:improvement-detection}
\begin{algorithmic}[1]
\REQUIRE Circuit $C$, List of detected CNOT chains $\mathcal{L}$, parameter mode $\in \{\texttt{conservative}, \texttt{always}\}$
\STATE $D_{\text{base}} \gets \textsc{Depth}(C)$
\FOR{each chain $ch \in \mathcal{L}$}
    \STATE $\mathcal{T} \gets$ generate candidate transformations for $ch$
    \STATE $\mathcal{R} \gets [\,]$ \Comment{stores results of candidates}
    \FOR{each $t \in \mathcal{T}$}
        \STATE $C' \gets \textsc{Apply}(C, t)$
        \STATE $D' \gets \textsc{Depth}(C', \texttt{scope} = 100)$
        \STATE append $(t, D')$ to $\mathcal{R}$
    \ENDFOR
    \IF{mode = \texttt{conservative}}
        \STATE choose $(t^{*}, D^{*}) \in \mathcal{R}$ s.t. $D^{*} < D_{\text{base}}$
        \IF{$t^{*}$ exists}
            \STATE $C \gets \textsc{Apply}(C, t^{*})$
            \STATE $D_{\text{base}} \gets D^{*}$
        \ENDIF
    \ELSIF{mode = \texttt{always}}
        \State choose $(t^{*}, D^{*}) \in \mathcal{R}$ with minimal $D^{*}$
        \STATE $C \gets \textsc{Apply}(C, t^{*})$
        \STATE $D_{\text{base}} \gets D^{*}$
    \ENDIF
\ENDFOR
\STATE \RETURN{} $C$
\end{algorithmic}
\end{algorithm}

\algnewcommand\AND{\textbf{and }}

\begin{algorithm}[h!]
\caption{DetectAndDecomposeChains: CX/CZ Chain Detection}\label{alg:detectChainsImproved}
\begin{algorithmic}[1]
\REQUIRE List of quantum instructions $L$
\STATE $i \leftarrow 0$
\WHILE{$i < \textit{length}(L)$}
    \STATE $l_i \leftarrow L[i]$
    \IF{\textit{isCX}($l_i$)}
        \STATE $C \leftarrow [l_i]$
        \STATE $j \leftarrow i + 1$
        \WHILE{$j < \textit{length}(L)$}
            \STATE $l_j \leftarrow L[j]$
            \IF{\textit{isCX}($l_j$) \AND \textit{isChainLike}($C[-1], l_j$)}
                \STATE \textit{append}($l_j, C$)
                \STATE $j \leftarrow j + 1$
            \ELSIF{\textit{isSingleQubitGate}($l_j$, \{$rx$, $ry$, $rz$\})}
                \STATE \textbf{break} \COMMENT{Chain broken by rotation}
            \ELSE
                \STATE \textbf{break} \COMMENT{Chain broken by non-chainlike CX}
            \ENDIF
        \ENDWHILE
        \IF{$\textit{length}($C$) > 1$}
            \STATE \textit{storeChain}($C$)
            \STATE \textit{decomposeChain}($C$)
            \STATE $i \leftarrow \text{\textit{startIndex}}(C)$
        \ELSE
            \STATE $i \leftarrow i + 1$
        \ENDIF
        \STATE \textbf{continue}
    \ELSIF{\textit{isCZ}($l_i$)}
        \STATE $C \leftarrow [l_i]$
        \STATE $j \leftarrow i + 1$
        \WHILE{$j < \textit{length}(L)$}
            \STATE $l_j \leftarrow L[j]$
            \IF{\textit{isCZ}($l_j$) \AND \textit{isChainLike}($C[-1], l_j$)}
                \STATE \textit{append}($l_j, C$)
                \STATE $j \leftarrow j + 1$
            \ELSIF{\textit{isSingleQubitGate}($l_j$, \{$rx$, $ry$, $rz$\})}
                \STATE \textbf{break}
            \ELSE
                \STATE \textbf{break}
            \ENDIF
        \ENDWHILE
        \IF{$\textit{length}($C$) > 1$}
            \STATE \textit{storeChain}($C$)
            \STATE \textit{decomposeChain}($C$)
            \STATE $i \leftarrow \text{\textit{startIndex}}(C)$
        \ELSE
            \STATE $i \leftarrow i + 1$
        \ENDIF
        \STATE \textbf{continue}
    \ELSE
        \STATE $i \leftarrow i + 1$
        \STATE \textbf{continue}
    \ENDIF
\ENDWHILE
\end{algorithmic}
\end{algorithm}

\subsection{Trade-off between circuit depth reduction identification and compilation speed}

Our approach explicitly balances the trade-off between compilation speed and optimization quality. In the default mode, the compiler analyzes whether a chain decomposition reduces circuit depth before applying it, ensuring that transformations contribute positively to overall circuit quality.

Alternatively, users may enable a mode by setting the optimization parameter to always apply. In this mode, the compiler bypasses the depth evaluation step and unconditionally decomposes any \texttt{CX} chain longer than five gates. Even though more chains will now be transformed, the compiler now no longer needs to calculate the circuit depth of the CNOT chain and the following 100 instructions to determine if circuit depth reduction has been achieved. Next to this, all CNOT gates in the chain are now transformed so they can no longer be part of subsequent CNOT chains, making the compiler skip over them faster. 
While this guarantees consistent application of the optimization and reduces compilation overhead, it may occasionally increase the circuit depth. This option is therefore well-suited to scenarios where runtime efficiency of the compiler is prioritized over strict depth minimization.

From an implementation perspective, we avoid a two-pass approach, where chains are first identified globally and then decomposed in a subsequent sweep. Instead, transformations are applied on the fly during traversal of the instruction list. This reduces computational overhead by eliminating the need for a second traversal. Furthermore, once a gate is incorporated into a chain, it is marked as processed, allowing subsequent passes to skip it entirely. This prevents redundant evaluations of gates already assigned to a chain and yields additional speedup.

\section{Evaluation}\label{sec:analysis}

The effectiveness of the proposed compiler optimizations can be evaluated by analyzing the impact of different GHZ state creation and chain implementations on circuit depth and gate count. The figures and discussions in following sections illustrate these improvements.

The compilations were performed on a standard laptop (12th Gen Intel(R) Core(TM) i7-1265U (1.80 GHz) CPU with 16GB RAM). On average, the benchmarks consisted of approximately 18,000 instructions and required about 15 seconds each to complete. The largest benchmark compiled contained 98,000 instructions and completed in 32 seconds. The longest compilation time observed was 82 seconds, corresponding to a quantum circuit with 52,000 instructions. This circuit, in particular, required more time due to the high density of CNOT gates forming potential CNOT chains. In such cases, the compiler frequently jumped back in the circuit, which increases the overall compilation time.

\subsection{Impact on circuit depth and gate count}
The GHZ state creation process shows significant variations in resulting depth and gate count across the different implementations.

\begin{figure*}[!tbp]
  \centering
  \begin{subfigure}[b]{0.25\textwidth}
    \includegraphics[width=\textwidth]{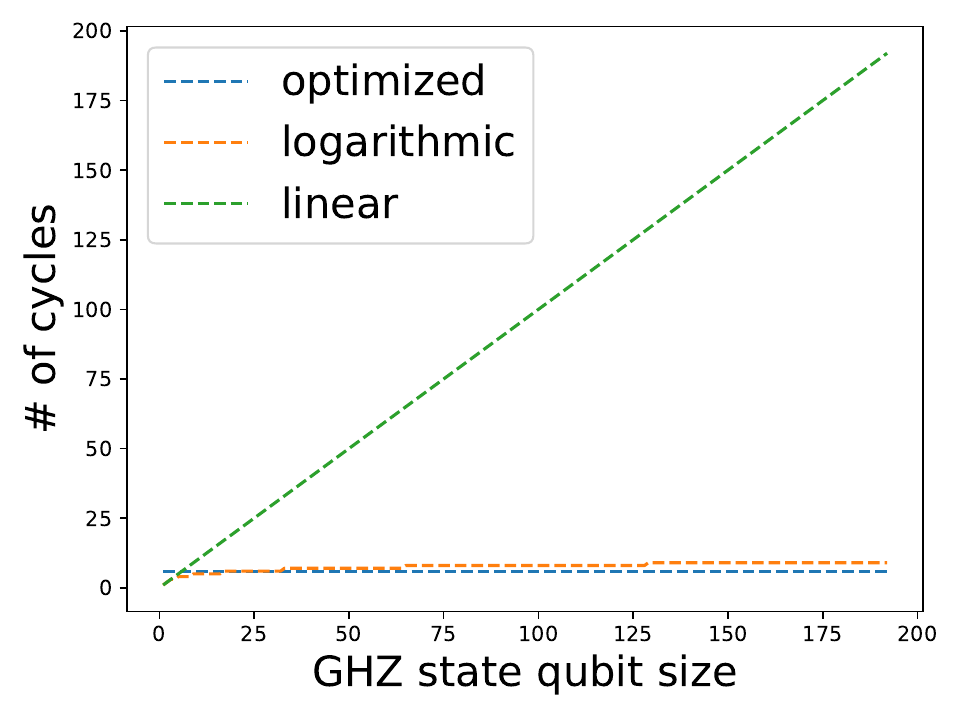}
    \caption{}\label{fig:GHZ_state_meas}
  \end{subfigure}
  \hfill
  \begin{subfigure}[b]{0.25\textwidth}
    \includegraphics[width=\textwidth]{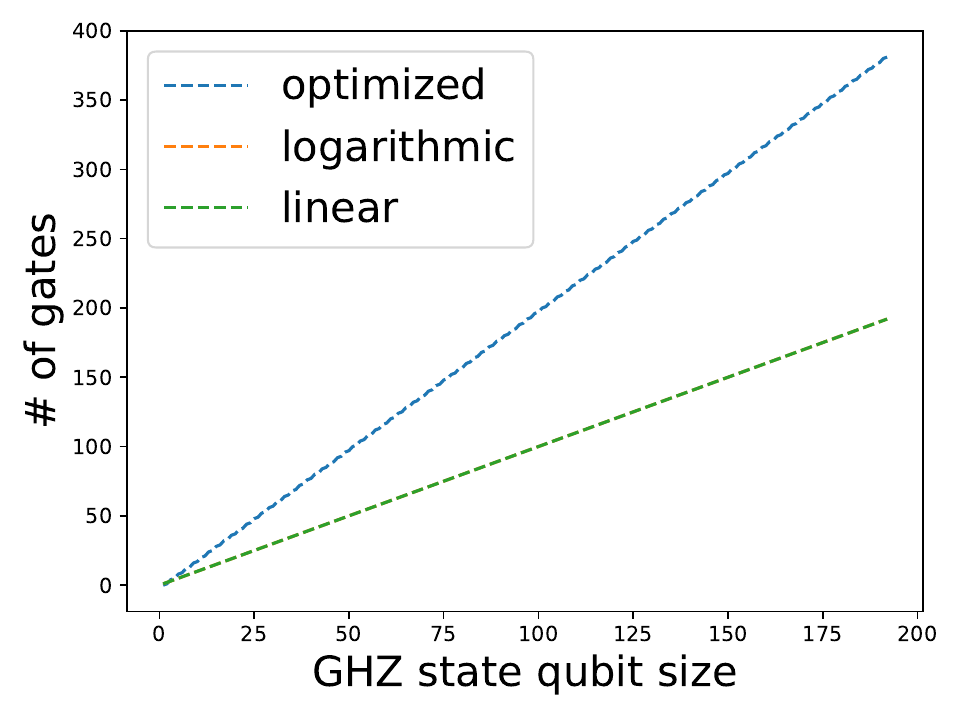}
    \caption{}
    \label{fig:gates_ghz}
  \end{subfigure}
  \hfill
  \begin{subfigure}[b]{0.25\textwidth}
    \includegraphics[width=\textwidth]{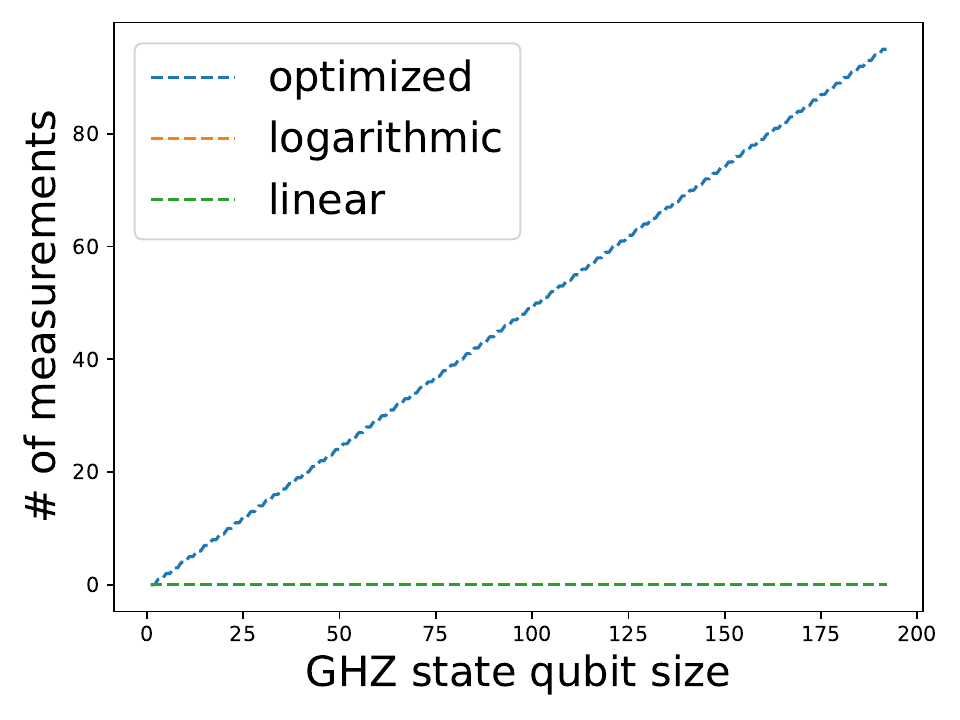}
    \caption{}
    \label{fig:ghz_state_depth}

  \end{subfigure}
  \caption{
  The impact of different GHZ state algorithms are illustrated.
  (a) Circuit depth: the original approach scales linearly, while the improved approach results in logarithmic depth and the most optimized approach results in constant depth.
(b) Gate count: All three versions exhibit linear scaling in gate count, with the optimized approach scaling more poorly.
(c) Measurement count: The original and improved versions both avoid measurements, resulting in zero measurement overhead, while the optimized approach exhibits linear scaling.
  }
\end{figure*}

As shown in Fig. \ref{fig:ghz_state_depth}, the original GHZ state creation method (green line) exhibits a linear growth in circuit depth as the number of qubits increases. This is a result of the sequential application of CNOT gates. The logarithmic version reduces circuit depth to logarithmic scaling by enabling a larger number of qubits to participate in simultaneous operations. The most optimized implementation further reduces depth to a constant level, achieving maximal parallelism at the cost of introducing additional measurement operations, as illustrated in Fig. \ref{fig:GHZ_state_meas}.

Gate count follows a similar trend, as illustrated in Fig. \ref{fig:gates_ghz}. The most parallel approach increases the number of gate operations significantly. The approaches thus show a trade-off between gate-amount, amount of measurements and circuit depth.


\begin{figure*}[!tbp]
  \centering
  \begin{subfigure}[b]{0.33\textwidth}
    \includegraphics[width=\textwidth]{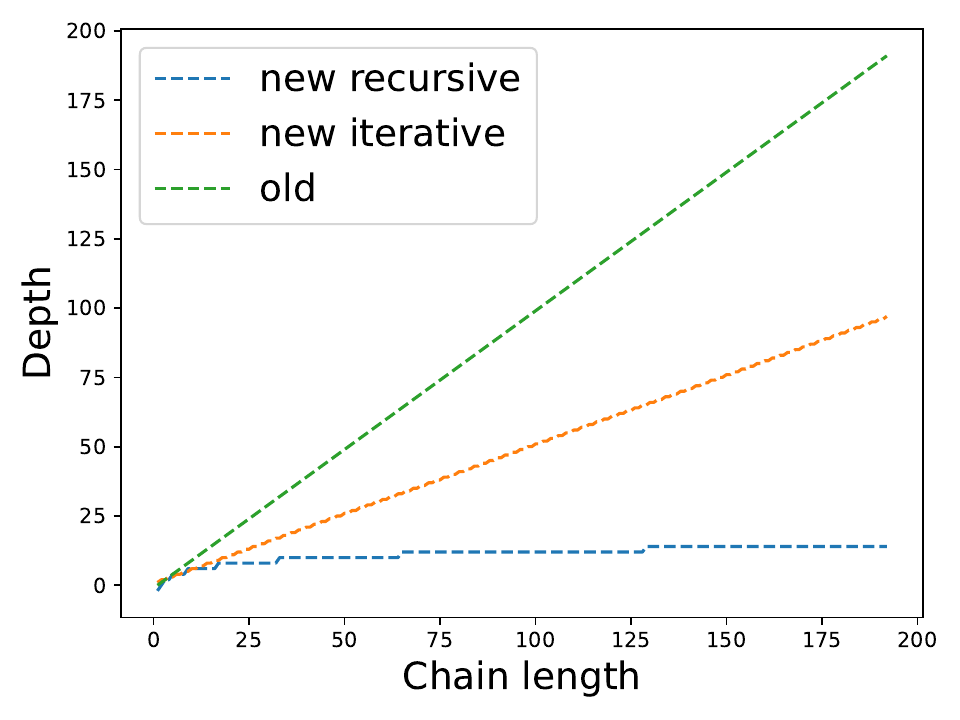}
    \caption{}\label{fig:cycles_chain}
  \end{subfigure}
  \begin{subfigure}[b]{0.33\textwidth}
    \includegraphics[width=\textwidth]{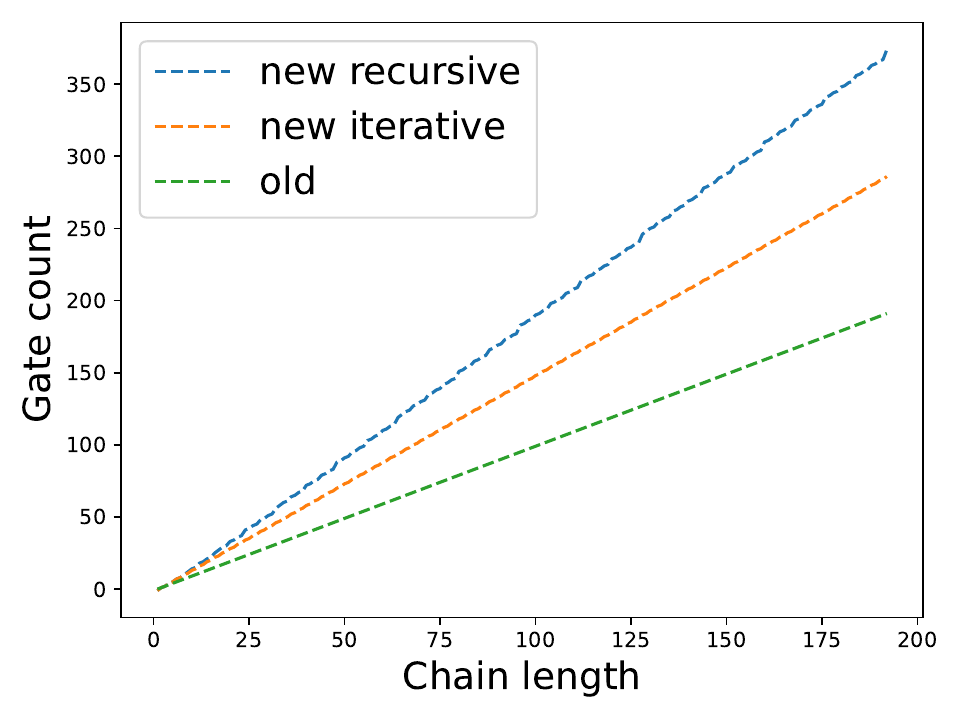}
    \caption{}
    \label{fig:CNOT_gates}
  \end{subfigure}
  \caption{
The theoretical potential of chain optimization is illustrated.
(a) Circuit depth comparison for three versions: the old and single-iteration versions scale linearly, with the latter achieving half the depth. The recursive version scales logarithmically.
(b) Gate count comparison: the recursive solution uses approximately twice as many gates as the original.}
\end{figure*}

The decomposition of CNOT chains exhibits substantial improvements in circuit depth as well, even tHough it shows an increase in gate count. As shown in Fig. \ref{fig:cycles_chain}, the traditional implementation of a CNOT chain (green line) results in a linearly increasing circuit depth. By applying a structured decomposition and allowing for partial parallelization, the optimized implementation (shown in orange) substantially reduces this depth. The recursive optimization further enhances performance by breaking down the chain into smaller segments that can be executed in parallel.

While these optimizations increase the number of gates (see Fig. \ref{fig:CNOT_gates}), the overall circuit depth is reduced to such an extent that the added gate overhead is a worthwhile trade-off in the context of noisy intermediate-scale quantum (NISQ) hardware.

\subsection{Accuracy and impact of CNOT Chain Detection}
We evaluated the effectiveness of our compiler on a suite of benchmark circuits to assess both, the accuracy of CNOT chain detection and the impact of deco0mposition on circuit depth. The results confirm that the detection algorithm reliably identifies CNOT chains across a variety of circuits. 

\begin{figure}
    \centering
    \includegraphics[width=0.7\linewidth]{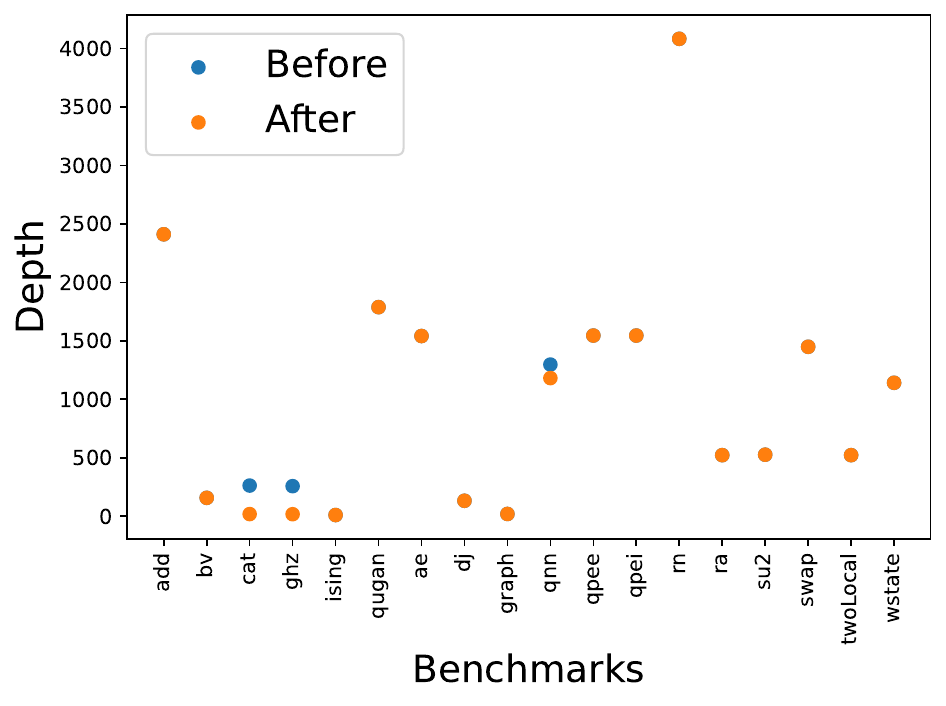}
    \caption{
    The graph compares benchmark circuit depths before (blue) and after (orange) using the compiler. Missing blue dots indicate that no change occurred. The results show that the compiler effectively identifies benchmarks that benefit from the chain reduction strategy.}
    \label{fig:algorithm_depth_cx_cz}
\end{figure}

As shown in Fig. \ref{fig:algorithm_depth_cx_cz}, the compiler applies decompositions selectively: a transformation is performed only when it leads to a reduction in circuit depth. When decomposition would instead increase or leave the depth unchanged, the compiler refrains from applying the transformation. 

These benchmarks highlight the compiler's ability to balance detection with compiler execution time, preserving or improving the circuit depth of the benchmarks.


\subsection{Depth reduction in VQEs}



\begin{figure*}
    \centering
    \includegraphics[width=1\linewidth]{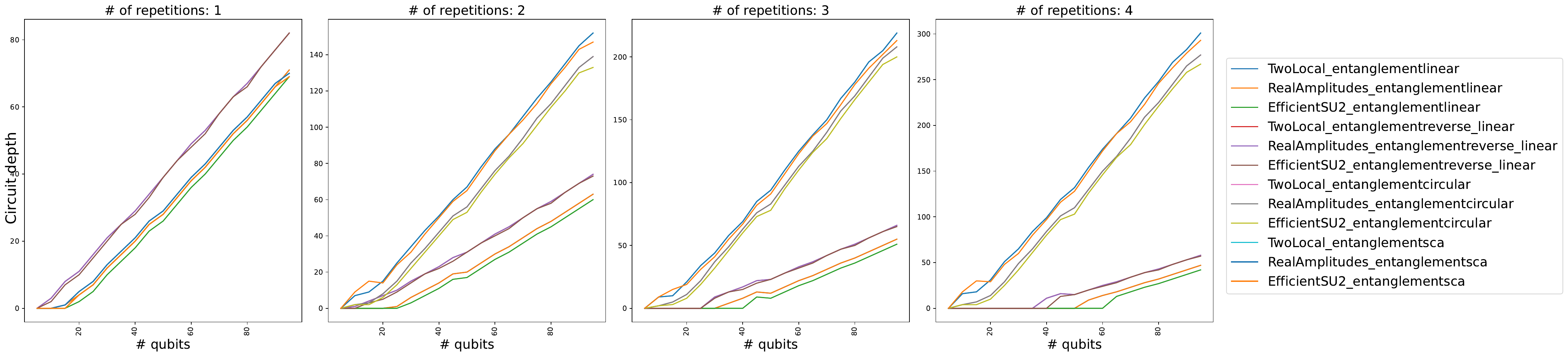}
    \caption{
    The plot shows VQEs that improved using our compiler, depicting the relative change in circuit depth before and after compilation (relative depth). Different benchmarks respond differently: some scale consistently with repetitions, while others show diminishing improvements as repetitions increase. The figure also demonstrates that the compiler can avoid applying transformations when unnecessary, as indicated by some algorithms maintaining a relative depth of 0 until the number of qubits exceeds a certain threshold.}
    \label{fig:VQE_always_good}
\end{figure*}

We evaluated our compiler on Variational Quantum Eigensolver (VQE) circuits across a range of ansätze and entanglement patterns (Fig. \ref{fig:VQE_always_good}). For qubit counts between $5$ and $95$, the compiler consistently reduces circuit depth while preserving logical equivalence. The observed depth reductions scale with circuit size, confirming the effectiveness and scalability of the optimization.

The impact of these optimizations depends strongly on circuit structure. Different ansätze and entanglement strategies (e.g., \texttt{circular}, \texttt{reverse\_linear}, \texttt{sca}) as well as circuit classes (e.g., \texttt{EfficientSU2}, \texttt{RealAmplitudes}, \texttt{TwoLocal}) respond differently to the applied transformations. Circuits with limited parallelism and identifiable substructures benefit the most, whereas already shallow or highly parallel circuits offer less opportunity for further depth reduction.

\begin{figure*}
    \centering
    \includegraphics[width=1\linewidth]{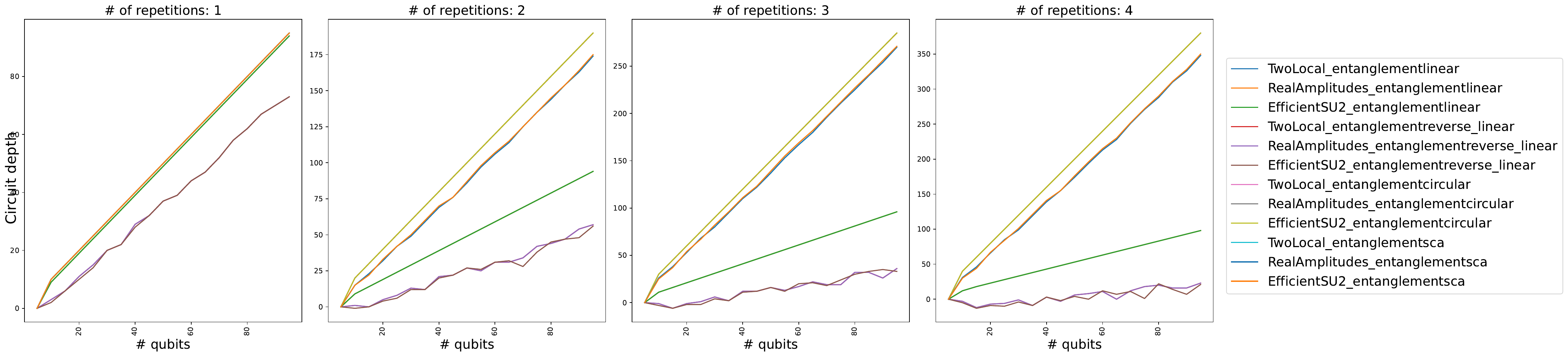}
    \caption{
    The plot shows VQEs that improved using our compiler, depicting the relative depth change before and after compilation. Here, the compiler was not restricted to apply transformations only when beneficial. The figure shows that some transformations actually increased circuit depth, highlighting the need for a compiler that can detect when improvements occur.
    }
    \label{fig:VQE_not_always_good}
\end{figure*}

Analysis of relative depth (i.e., the difference between pre- and post-compilation depth) further illustrates this behavior. In VQEs with multiple repetitions (Fig. \ref{fig:VQE_not_always_good}), relative depth reductions diminish and may even become negative for small qubit counts. This occurs because repeated layers already introduce parallelism, leaving fewer opportunities for additional optimization.

A key strength of the compiler is its selective application of transformations. As shown in Fig. \ref{fig:VQE_always_good}, no negative relative depth is observed, indicating that the compiler successfully avoids harmful transformations. In contrast, when transformations are applied unconditionally (Fig. \ref{fig:VQE_not_always_good}), some circuits exhibit increased depth, highlighting the importance of the compiler’s decision-making mechanism.

\section{Conclusion}\label{sec:conclusion}
In this work, we introduced a compiler-based approach to optimize quantum circuits by automatically detecting and transforming common subroutines: GHZ state creation and CNOT/CZ chains. Our compiler leverages structured decompositions, recursive strategies, and user-defined preferences to reduce circuit depth while maintaining logical equivalence. These transformations offer scalable improvements particularly well-suited for near-term quantum devices, where reducing depth is critical for minimizing decoherence and noise-induced errors.

Our analysis shows that GHZ state creation, which traditionally suffers from linearly increasing depth, can be optimized to logarithmic or even constant depth through our compiler transformations. Similarly, our recursive decomposition strategy for CNOT chains reduces depth logarithmically, at the cost of a doubled gate count of the CNOT chain in the worst case, a trade-off that aligns well with the priorities of NISQ-era quantum computing. Importantly, the compiler incorporates a built-in safeguard: it evaluates whether each decomposition actually lowers depth and only applies the transformation if it does. As a result, the compiler is guaranteed never to increase circuit depth.


We also evaluated our approach in the context of the Variational Quantum Eigensolver (VQE). Here, our compiler consistently reduced circuit depth, particularly for circuits with a single layer of parameterized gates. However, for some VQE circuits with repeated layers (i.e., \texttt{num\_reps} $> 1$), depth improvements diminish. This behavior stems from the fact that repeated layers often already contain parallel gate structures, leaving fewer opportunities for additional parallelization. 

Finally, we observed that the effectiveness of our compiler is algorithm-dependent. Different ansätze and entanglement strategies exhibit varying levels of optimization potential. Circuits with deep or sequential gate structures benefit the most, whereas highly parallel templates show less gains.

In conclusion, our compiler establishes itself as a reliable tool for quantum circuit optimization. It is targeted for circuit depth reduction of inherently sequential algorithms. Therefore, the largest circuit depth reduction is also seen in inherently sequential algorithms, while the depth of inherently parallel algorithms remains the same. 

This makes it particularly well-suited for the NISQ era, where every layer of depth matters for mitigating decoherence and noise. Looking ahead, the compiler offers a solid foundation for further advances.
The compiler could be extended by finding other subroutines that are inherently sequential and transforming them into parallel counterparts. Another potential future work could be to take either connectivity or noise values of the hardware into account while transforming a circuit. 



\section{Acknowledgements}
We gratefully acknowledge support from the joint research program “Modular quantum computers” by Fujitsu Limited and Delft University of Technology, co-funded by the Netherlands Enterprise Agency under project number PPS2007.

\bibliographystyle{IEEEtran}
\bibliography{report.bib}



\end{document}